\documentstyle[12pt,a4]{article}
\def\theequation{\arabic{equation}}

\renewcommand{\thefootnote}{\fnsymbol{footnote}}
\newcommand{\r}[1]{(\ref{#1})}
\begin{document}
\thispagestyle{empty}

\begin{center}
{\large{\bf AREA-PRESERVING ALGEBRA STRUCTURE
OF TWO-DIMENSIONAL SUPERGRAVITY}} \vspace{1.5cm}\\

{\large D.R. Karakhanyan\footnote{E-MAIL:karakhian@vxc.
erphy.am}}\vspace{0.5cm}

{\it Theoretical Physics Department,}\\
{\it Yerevan Physics Institute}\\
{\it Alikhanyan Br. st.2, Yerevan, 375036 Armenia }
\end{center}
\bigskip
\bigskip
\bigskip
\bigskip
\begin{abstract}

The effective action for 2d-gravity in Weyl-invariant regularization
is extended to supersymmetric case. The super area-preserving invariance
and cocyclic properties under general supergravitational transformations
of the last action is shown.

 \end{abstract}

\vfill
\setcounter{page}0
\renewcommand{\thefootnote}{\arabic{footnote}}
\setcounter{footnote}0
\renewcommand{\theequation}{\arabic{section}.\arabic{equation}}
\setcounter{equation}{0}
\newpage

\newcommand{\ra}{\rightarrow}

\pagestyle{plain}

\section{\bf Introduction.}
\setcounter{equation}{0}

	The Weyl symmetry in the string theory is a fundamental
principle which governs the string dynamics \cite{Pol}.

	The super Weyl transformations, which firstly had been
considered by Howe \cite{how} are useful for investigation of the
superspace constraints in the supergravity theories \cite {GSW}.

	Even being spoiled by quantum corrections the Weyl symmetry
of classical theory implies the cocyclic property of the effective action.
It is so restrictive that fixes action up to numerical constants
\cite {RRD}. This form of the effective action causes an independent
interest as the theory of matter field in (super) Weyl
invariant (super) gravitational background \cite{Jac}.

        In previous work \cite{KMM} we have studied transformation
properties of the effective action of 2d-gravity under Weyl rescaling
of metric.
The well known non-local functional of effective action of 2d-gravity is

\begin{equation}
\label{1}
       W[g]=\int R\frac{1}{\Box}R
\end{equation}

The Weyl-coboundary of \r{1}
\begin{equation}
\label{2}
       W[e^\sigma g]-W[g]\equiv S_L (\sigma , g)
\end{equation}
coincides with the famous Liouville action
\begin{equation}
\label{3}
S_L (\sigma,g)=\int d^2 x\sqrt g (\frac{1}{2}g^{\alpha\beta}
\partial_\alpha\sigma\partial_\beta\sigma + R\sigma)
\end{equation}
The action \r{3} satisfies to cocyclic condition
\begin{equation}
\label{4}
S_L(\sigma_1+\sigma_2,g)-S_L(\sigma_1,e^{\sigma_2}g)-S_L(\sigma_2,
g)=0
\end{equation}
which follows from \r{2}.The relation \r{4} allow us to consider
\r{3} as the anomaly arising from mesure's transformation in the
bosonic string theory:

\begin{equation}
\label{5}
D_{e^\sigma g}X^\mu=e^{-D S_L(\sigma,g)}D_gX^\mu
\end{equation}
where $D$ is dimension of target space.
The effective action of 2d-gravity, computed in the Weyl
invariant regularization is simply functional \r{1}, depending on the
Weyl invariant combination $g_{\alpha\beta}/\sqrt g$
\begin{equation}
\label{6}
\tilde W[g_{\alpha\beta}]=W\left[\frac{g_{\alpha\beta}}{\sqrt g}\right]=
\int (R(g)-\Box\log\sqrt g)\sqrt g\frac{1}{\sqrt g\Box}(R(g)-
\Box\log\sqrt g)\sqrt g.
\end{equation}
and differ from \r{1} by the local counterterm:
\begin{equation}
\label{7}
\tilde W[g]=W[g]+S_L(\sigma,g)|_{\sigma=-\log\sqrt g}
\end{equation}
However the Weyl invariant action \r{7} is not scalar under general
coordinate transformation $Diff_2$ now, it is invariant only under area-
preserving $SDiff_2$ subgroup of $Diff_2$. So under
shifting \r{7} we trade $Diff_2$ for $Weyl\times SDiff_2$ and
corresponding functional mesure is not diffeomorphism invariant.
\begin{equation}
\label{8}
\tilde D_{f^\ast g}X^\mu=e^{-dS(f,g)}\tilde D_gX^\mu.
\end{equation}
Here we used the notations
$$
S(f,g)\equiv S_L(\sigma,g)|_{\sigma=\log\Delta^F_x},\ \
\Delta^F_x=\det\|\frac{\partial F^\beta}{\partial x^\alpha}\|
$$
and
$F^\alpha$ is inverse to $f^\alpha$:
$F^\pm(f^+,f^-)=x^\pm$

\section{\bf The super Liouville action}
\setcounter{equation}{0}

The extension of procedure, described above, to case of the 2d
supergravity is transparent.
The super Liouville action consist from two parts: kinetic and
topological.
\begin{equation}
\label{9}
S_{S-L}(\sigma,\psi;e^a_\alpha,
\chi_\alpha)=S_0+S_1
\end{equation}
The kinetic part of \r{9} is well known Neveu-Schwarz string
action:
\begin{equation}
\label{10}
S_0=\int d^2 x
e(g^{\alpha\beta}\partial_\alpha\sigma\partial_\beta-i\bar\psi\gamma^
\beta\gamma^\alpha\partial_\alpha\psi+2\bar\chi_\alpha\gamma^\beta\gamma
^\alpha\psi\partial_\beta\sigma+\frac{1}{2}\bar\psi\psi\bar\chi_\alpha
\gamma^\beta\gamma^\alpha\chi_\beta)
\end{equation}
and topological part is
\begin{equation}
\label{11}
S_1=\int d^2 x\epsilon^{\alpha\beta}(\omega_\alpha(e)+\frac{1}{2}\bar
\chi_\alpha\gamma_3\gamma^\mu\chi_\mu)(\partial_\beta\sigma+\bar\chi_\nu
\gamma_\beta\gamma^\nu\psi)
\end{equation}
The $S_0$ and $S_1$ are separately invariant under local
supersymmetry transformations. Their sum satisfies to the cocyclic
condition:
\begin{eqnarray}
\label{12}
&&S_{S-L}(\sigma_1+\sigma_2,\psi_1;e^a_\alpha,\chi_\alpha)=
S_{S-L}(\sigma_2,\psi_2;e^a_
\alpha,\chi_\alpha)+\nonumber\\
&&\hspace{2cm}+S_{S-L}(\sigma_1,
e^{-\frac{\sigma_2}{4}}\psi_1;e^{\frac{\sigma_2}{2}}e^a_\alpha,e^{\frac{
\sigma_2}{4}}(\chi_\alpha+\gamma_\alpha\psi_2)
\end{eqnarray}
In agreement with general analisis of the cocycles of the Weyl
group \cite{RRD} we note that expression \r{9} is polynomial
on parameters of the super Weyl group ($ \sigma$ and $\psi$) of
degree no more that two (dimension of the space-time).
The maximal order terms on $\sigma$ and $\psi$ i.e.
$S_0$ are invariant under super Weyl transformations:
\begin{eqnarray}
\label{13}
\psi &\rightarrow & e^{-\frac{\sigma}{4}}\psi \nonumber\\
e^a_\alpha &\rightarrow & e^{\frac{\sigma}{2}}e^a_\alpha\\
\chi_\alpha & \rightarrow & e^{\frac{\sigma}{2}}\chi+\gamma_\alpha
\eta\nonumber
\end{eqnarray}
while the low-order part $S_1$ is not in complete agreement with
general analisis of the cocycles of the Weyl group \cite{RRD}.

	The superfield formulation is rather simple.
In terms of superfield
$\varphi=\sigma+\bar\theta\psi+\bar\psi\theta+B\bar\theta\theta$
expression \r{9} takes the form
\begin{equation}
\label{14}
S_{S-L}=\int d^2 xd^2\theta E(\frac{1}{2}D\varphi\bar D\varphi
+\Re\varphi)
\end{equation}
where $\Re$ is supercurvature.

The expression \r{14} is super Weyl coboundary of the non-local
effective action
\begin{equation}
\label{15}
W[E]=\int \Re\frac{1}{D\bar D}\Re.
\end{equation}
It has general form
\begin{equation}
\label{16}
(Anomaly)\Delta^{-1}(Anomaly)
\end{equation}
where $\Delta$ is (super)Weyl invariant differential operator of
order $d$ (space-time dimension).
To obtain super Weyl invariant expression one needs to replace
zweibein and gravitino by super Weyl invariant combinations
\begin{equation}
\label{17}
e^a_\alpha\rightarrow\frac{e^a_\alpha}{\sqrt g},\quad
\chi_\alpha\rightarrow
e^{-\frac{1}{4}}\gamma^\beta\gamma_\alpha\chi_\beta
\end{equation}
To define, which part of two-dimensional superdiffeomorphysm group
is compatible with super Weyl symmetry, it should to be find local
supersymmetric and general coordinate transformations whose
preserve gauge conditions:
\begin{equation}
\label{18}
\det\|e^a_\alpha\|=1,\hfil \gamma^\alpha\chi_\alpha=0
\end{equation}
The answer is following: it is diffeomorphysms with divergentless
vector field and local supersymmetry with the parameter, which
satisfies constraint
\begin{equation}
\label{19}
\gamma^\alpha(\partial_\alpha-\frac{1}{2}\omega_\alpha(e))\epsilon
(x)=0
\end{equation}
Let us notice that the commutator of such transformations gives
rise to area-preserving diffeomorphysm.
This is quite natural, because that transformations form maximal
subalgebra of two-dimensional superdiffeomorphisms.
This subgroup is defined by the symplectic form
\begin{equation}
\label{20}
\Omega=pdq-\theta^Ad\theta^A
\end{equation}
in phase superplane. The corresponding Poisson braket is
\begin{equation}
\label{21}
\{ A,B\} =\epsilon^{\alpha\beta}\partial_\alpha A\partial_
\beta B+\delta^{AB}\frac{\partial^LA}{\partial\theta^A}
\frac{\partial^RB}{\partial\theta^B}
\end{equation}
If we choose super Weyl invariant regularization for functional
mesure, then under general superdiffeomorphysms it is transformed
according to rule
\begin{equation}
\label{22}
\tilde D_{F\ast E}\Phi=e^{S(F,E)}\tilde D_E\Phi
\end{equation}
Here $F$ is superdiffeomorphysm, acting on supervielbein $E$,
constrained by unit determinant and $\Phi$ is matter superfield.

\section{\bf Conclusion and outlook}

	As we seen, all regularities of ordinary 2d-gravity
can be transformed to the case of supergravity in natural way.

	It seems real to do this in four and higher dimensions.

	This approach can be applied to w-gravity. Because the 2d
supergravity in many aspects seems like to w-gravity \cite{OSSN} it is
reasonable to try find nonchiral expression for the effective action of
$w_3$-gravity in flat and general ordinary- and $w_3$- gravitational
background.

\section{\bf Acknowledgments}

	I would like to thank F. David, P. West, I. Kostov, R. Manvelyan
and R. Mkrtchyan for helpful discussions. Finally, I would like to thank
the CEA Saclay for hospitality while the part of this work was done.

This work is supported in part by the
NATO linkage grant  LG 9303057, the
grant 211-5291 YPI of the
German Bundesministerium f\"ur Forshung und Technologie, Federal
Republic of Germany and the grant INTAS -- 2858


\begin{thebibliography}{99}
\bibitem{Pol} A.M.Polyakov, Phys.Lett.{\bf B 103} (1981) 893\\
A.M.Polyakov, Mod.Phys.Lett.{\bf A 2} (1987) 893
\bibitem{how} P.S. Howe J.Phys. {\bf A}: Math.Gen. 12 (1979) 393
\bibitem{GSW} S. James Gates Jr., K.S. Stelle, P.C. West, Nucl. Phys.
{\bf B} 169 (1980) 347
\bibitem{RRD} D.R. Karakhanyan, R.P. Manvelyan, R. L. Mkrtchyan,
hep-th/9411068 (submitted to Nucl. Phys.)
\bibitem{Jac} R. Jackiw, {\em Another View on Massless Matter-Gravity
Fields in Two-Dimension}, preprint MIT-CTP-2377\\
D. Candemi, R. Jakiv, B. Zwiebach, {\em Physical States in Matter-Coupled
Dilaton Gravity}, preprint UCLA/95/TEP/16, MIT-CTP-2436\\
G. Amelino-Camelia, D. Seminara, preprint MIT-CTP-2429, hep-th/9505136\\
G. Amelino-Camelia, D. Seminara, preprint MIT-CTP-2443, hep-th/9506128
\bibitem{KMM}D.R. Karakhanyan, R.P. Manvelyan and R.L. Mkrtchyan,
Phys. Lett. {\bf B 329} (1994)185
\bibitem{OSSN} H. Ooguri, K. Schoutens, A. Sevrin, P. van Nieuwenhuizen
Comm. Math. Phys. {\bf 145} (1992) 515.

\end{thebibliography}
\end{document}